\documentclass[11pt,twoside]{article}

\usepackage{asp2006}
\usepackage{epsf}
\usepackage{psfig}
\usepackage{lscape}

\begin{document}
\title{Binary Star Detection and Parameter Estimation in the RAVE Survey}
\author{Gal Matijevi\v c\altaffilmark{1}, Toma\v z Zwitter\altaffilmark{1}, and the RAVE Collaboration}
\affil{$^1$ University of Ljubljana, Faculty of Mathematics and Physics, Jadranska 19, 1000 Ljubljana, Slovenia}

\begin{abstract}
Although primarily aimed at the galactic archeology and evolution, automated all-sky spectroscopic surveys (RAVE, SDSS) are also a valuable source for the binary star research community. Identification of double-lined spectra is easy and it is not limited by the rare occurrences of eclipses. When the spectrum is properly classified, its atmospheric parameters can be calculated by comparing the spectrum with the best fit atmosphere model. We present the analysis of the binary stars from the sample of roughly 250.000 RAVE survey spectra. The classification and binary discovery method is based on the correlation function analysis. The comparison of these spectra with the model shows that it is possible to estimate the essential atmospheric parameters relatively well. Large number of such estimates and the fact that RAVE consists of a magnitude selected sample without any color cuts makes it suitable for a binary star population study.
\end{abstract}

\section{Introduction}

The standard way of solving binary star systems is through the use of sophisticated models, the Wilson-Devinney code for example. This approach is very appealing because it is possible to precisely recover a number of system's parameters. The disadvantage is in the fact that for a reliable solution one needs a large set of photometric as well as spectroscopic observations. The latter only serve as a mean for measuring radial velocities, usually discarding the abundance of other information carried by spectra. It has been proven many times on spectra of single stars \citep[e.g.~][to name just a few]{bailer-jones1997,katz1998,decin2004,ocvirk2006,koleva2009} that it is possible to estimate some of the astrophysical parameters (e.g.~temperature, gravity, and metallicity) by fitting the observed spectrum with the stellar atmosphere model. The same can be done in case of binary stars. This approach might be particularly useful in the pipelines of spectroscopic surveys, where parameters of large amount of observed objects have to be extracted automatically.

In the following section we will briefly review the RAVE survey itself and present the results of estimation of the most significant parameters from the spectra of binary stars in the RAVE survey. Before that we will also deal with another problem that has to be addressed when working with large datasets of unknown objects, the discovery of binary stars. Because it is not known in advance if some spectrum belongs to a binary star or some other type of peculiar object, an automatic method has to be used to properly classify all spectra.

\section{RAVE Survey}

\textit{RAdial Velocity Experiment} is a spectroscopic all-sky survey, primarily targeted at measuring stellar radial velocities. It is based on the $1.2\,\mathrm{m}$ UK Schmidt telescope. From the start in 2003, the fiber-fed  spectrograph gathered almost 400.000 medium resolution $(R\sim 7500)$ spectra in the spectral range from $840\,\mathrm{nm}$ to $880\,\mathrm{nm}$. Part of the measured data with the parameters for single stars is already available through first and second data releases \citep{steinmetz2006,zwitter2008}.

The spectral range was chosen because it includes a strong Ca II triplet, a few Paschen lines and is also rich in other metallic lines, enabling precise radial velocity measurement as well as other atmospheric parameters.

The sample of stars, selected for observation, is completely unbiased. The only criterion for inclusion in the catalog is an $\mathit{I}$ magnitude in the range from 9 to 12. The sample covers different stellar populations and includes significant amount of dwarfs as well as giants but also binary and peculiar stars. Most of the objects are observed once, only about ten percent of the observing time is dedicated to the repeated observation to check for consistency.

\section{Classification}

Historically, spectral types were assigned to stars by visual inspection. Nowadays, it seems almost impossible to rely on such methods, especially considering the amount of data that has to be processed. In order to automate this process, different methods were developed. Especially tempting are the methods that use artificial neural networks as a classification core since they are, once properly set, extremely fast in comparison to other methods. Unfortunately, tests made on RAVE spectra proved that such methods have trouble finding spectra with systematic errors, so we used the analysis of the correlation function (CF) between the observed spectrum and a fixed template model spectrum. Inspection of different regions of the CF give information about the type of the spectrum and also about the presence of noise or technical defects. The classification procedure is as follows:

\begin{itemize}
\item Firstly, the minimal and maximal values of the normalized spectrum are checked. If any of these two are completely out of usual limits, this give a strong indication of an extremely noisy or a corrupt spectrum.

\item Next step is the calculation of the CF, which is then smoothed and compared to the original. If the original is rough, the difference between the two will be significant, again giving the clue about noisy spectrum.

\item Several properties of the CF are then checked. Double peak as well as significant asymmetry of the peak indicate a binary spectrum. Uneven CF's wings are related to the unwanted oscillations in the continuum. Relatively low and wide peaks belong to the spectra of hot stars.

\item Next, simple analysis of the Ca II triplet lines is done. If they are unusually wide, the spectrum might belong to either binary, hot, or emission star.

\item Finally, if none of the above criteria is met, the spectrum is compared to the best model. Bad fit at a given S/N ratio might indicate non-regular object.
\end{itemize}

\begin{figure}[!ht]
\plotone{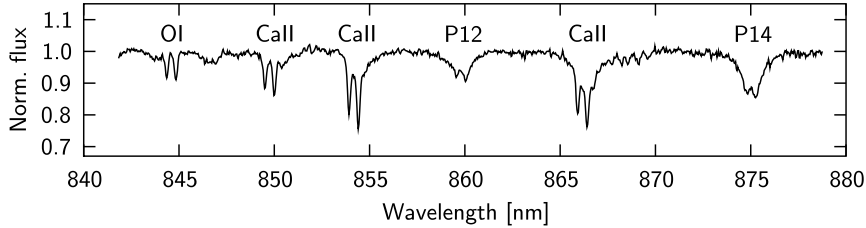}
\caption{Example of a high S/N double-lined spectrum produced by the RAVE survey. Split lines are clearly visible, especially all three Ca II lines and H lines from the Paschen series.}
\end{figure}

In the sample of $\sim 250.000$ spectra around one percent are declared as binary by the above method. Among the discovered binaries are only the ones with evident double lines and similar flux ratios (Figure 1). Interestingly though, only a few percent of the confirmed binaries are also declared as binary in the SIMBAD database, meaning it is possible to efficiently discover new binary stars.

\section{Parameter estimation}

When spectra are properly classified as binary, the parameters can be extracted by comparing the binary spectrum to a set of model spectra. This is done by first defining a criterion for goodness of fit (usually just the square of the difference between the observed and the model spectrum, summed through the whole spectral range) and then constructing the binary spectrum with an optimal fit to the observed one.

A set of parameters that are recoverable in binary spectra of the RAVE survey includes effective temperatures of both components, their surface gravities, chemical composition, radial velocities and flux ratio. The effect of line broadening by stellar rotation is typically too weak to be measurable at a given resolution. In case of two observations of the same object it is also possible to measure a mass ratio and a center-of-mass radial velocity of the binary system.

The fitting process was done in two stages. For each spectrum the initial Doppler shifts and flux ratio values were estimated using TODCOR method \citep{zucker1994}. Along with the initial values for other parameters, the best fitting model from a library of synthetic spectra \citep{munari2005} was then found using a simplex method.
Statistics on individual recovered parameters are shown in Figure 2. Roughly estimated errors of parameters are equal to $\sigma_{Teff}=500\,\mathrm{K},\  \sigma_{logg}=0.5\,\mathrm{dex},\ \sigma_{M/H}=0.2\,\mathrm{dex}$ and $\sigma_{L1/L2}=0.2$. 

\begin{figure}[!ht]
\plotone{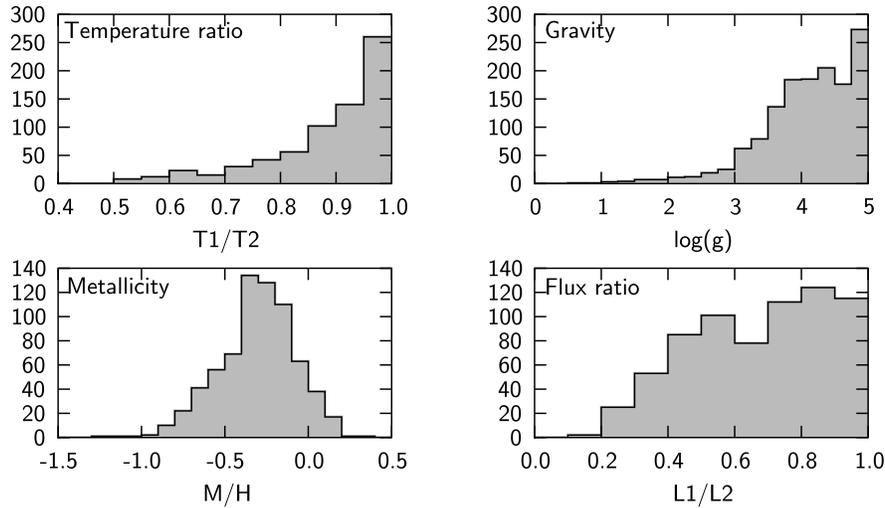}
\caption{Estimated parameters of binary star spectra in the RAVE survey.}
\end{figure}

Although the results seem reasonable, there are a few problems with using this method. The errors of parameters are hard to calculate, since the method provides only the point of minimum and not the surrounding area. The method is also relatively sensitive to initial conditions which makes it impossible to get sufficiently reliable result using a single starting point. Very recent tests using a nested sampling method \citep{feroz2008} proved this method gives more reliable results without the aforementioned problems and without significantly longer calculation time.

\section{Conclusion}

The described classification and parameter estimation methods gave satisfactory results to some extent, but there is still room for improvement. Other methods like the mentioned nested sampling seem very promising for such tasks. Also, the planned use of stellar evolution models will additionally constrain the errors of calculated parameters, yielding better results.
 
The forthcoming missions like Gaia and the Hermes survey, will provide even more data with higher resolution. The need for truly automatic processing pipelines will be even grater, so it is important to gain as much experience as possible until then.


\begin{thebibliography}
\bibitem[Bailer-Jones et al., 1997]{bailer-jones1997}
Bailer-Jones, C.~A.~L., Irwin, M., Gilmore, G. et al., 1997, \mnras, 292, 157
\bibitem[Decin et al., 2004]{decin2004}
Decin, L., Shkedy, Z., Molenberghs, G. et al., 2004, \aap, 421, 281
\bibitem[Feroz and Hobson, 2008]{feroz2008}
Feroz, F. and Hobson, M.~P., 2008, \mnras, 384, 449
\bibitem[Katz et al., 1998]{katz1998}
Katz, D., Soubiran, C., Cayrel, R. et al., 1998,\aap, 338, 151
\bibitem[Koleva et al., 2009]{koleva2009}
Koleva, M., Prugniel, P., Bouchard, A. et al., 2009, \aap, 501, 1269
\bibitem[Munari et al., 2005]{munari2005}
Munari, U., Sordo, R., Castelli, F. et al., 2005, \aap, 442, 1127
\bibitem[Ocvirk et al., 2006]{ocvirk2006}
Ocvirk, P., Pichon, C., Lanc\c on, A. et al., 2006, MNRAS, 365, 46
\bibitem[Steinmetz et al., 2006]{steinmetz2006}
Steinmetz, M., Zwitter, T., Siebert, A. et al., 2006, \aj, 132, 1645
\bibitem[Zucker and Mazeh, 1994]{zucker1994}
Zucker, S.~and Mazeh T., 1994, \apj, 420, 806
\bibitem[Zwitter et al., 2008]{zwitter2008}
Zwitter, T., Siebert, A., Munari, U. et al., 2008, \aj, 136, 421
\end{thebibliography}
\end{document}